\documentclass[12pt,preprint]{aastex}

\def\sun{\odot}

\def\icarus{Icarus}
\usepackage{epsfig}
\begin{document}
\title{Overstable Librations can account for the Paucity of Mean Motion Resonances among Exoplanet Pairs}

\shorttitle{A Simple Model for Orbital Resonances}

\author{Peter Goldreich\altaffilmark{1,2} and Hilke E. Schlichting\altaffilmark{3}}
\altaffiltext{1} {California Institute of Technology, MC 150-21, Pasadena, CA 91125, USA}
\altaffiltext{2}{Institute for Advanced Study, Princeton, NJ 08540, USA), AB(Institute for Advanced Study, Princeton, NJ 08540, USA}
\altaffiltext{3} {Massachusetts Institute of Technology, 77 Massachusetts Avenue, Cambridge, MA 02139-4307, USA}
\email{hilke@mit.edu}

\begin{abstract} 
We assess the multi-planet systems discovered by the Kepler satellite in terms of current ideas about orbital migration and eccentricity damping due to planet-disk interactions. Our primary focus is on first-order mean motion resonances, which we investigate analytically to lowest order in eccentricity.  Only a few percent of planet pairs are in close proximity to a resonance. However, predicted migration rates (parameterized by $\tau_n=n/{|\dot n|}$) imply that during convergent migration most planets would have been captured into first order resonances.   Eccentricity damping (parameterized by $\tau_e=e/{|\dot e|}$) offers a plausible resolution.  Estimates suggest $\tau_e/\tau_n\sim (h/a)^2\sim 10^{-2}$, where $h/a$ is the ratio of disk thickness to radius.  Together, eccentricity damping and orbital migration give rise to an equilibrium eccentricity, $e_{eq}\sim(\tau_e/\tau_n)^{1/2}$.  Capture is permanent provided $e_{eq}\lesssim \mu^{1/3}$, where $\mu$ denotes the planet to star mass ratio.  But for $e_{eq}\gtrsim \mu^{1/3}$,  capture is only temporary because librations around equilibrium are overstable and lead to passage through resonance on timescale $\tau_e$.  Most Kepler planet pairs have  $e_{eq}>\mu^{1/3}$.  Since $\tau_n\gg \tau_e$ is the timescale for migration between neighboring resonances, only a modest percentage of pairs end up trapped in resonances after the disk disappears.  Thus the paucity of resonances among Kepler pairs should not be taken as evidence for in situ planet formation or the disruptive effects of disk turbulence.  

Planet pairs close to a mean motion resonance typically exhibit period ratios 1-2\% larger than those for exact resonance.  The direction of this shift undoubtedly reflects the same
asymmetry that requires convergent migration for resonance capture.  Permanent resonance capture at these separations from exact resonance would demand $\mu (\tau_n/\tau_e)^{1/2}\gtrsim 0.01$, a value that estimates of $\mu$ from transit data and $(\tau_e/\tau_n)^{1/2}$ from theory are insufficient to match.  Plausible alternatives involve eccentricity damping during or after disk dispersal.

The overstability referred to above has applications beyond those considered in this investigation. It was discovered numerically by \citet{MW08} in their study of the tidal evolution of SaturnsÕ satellites.
\end{abstract}


\section{INTRODUCTION}
\begin{figure} [htp]
\centerline{\epsfig{file=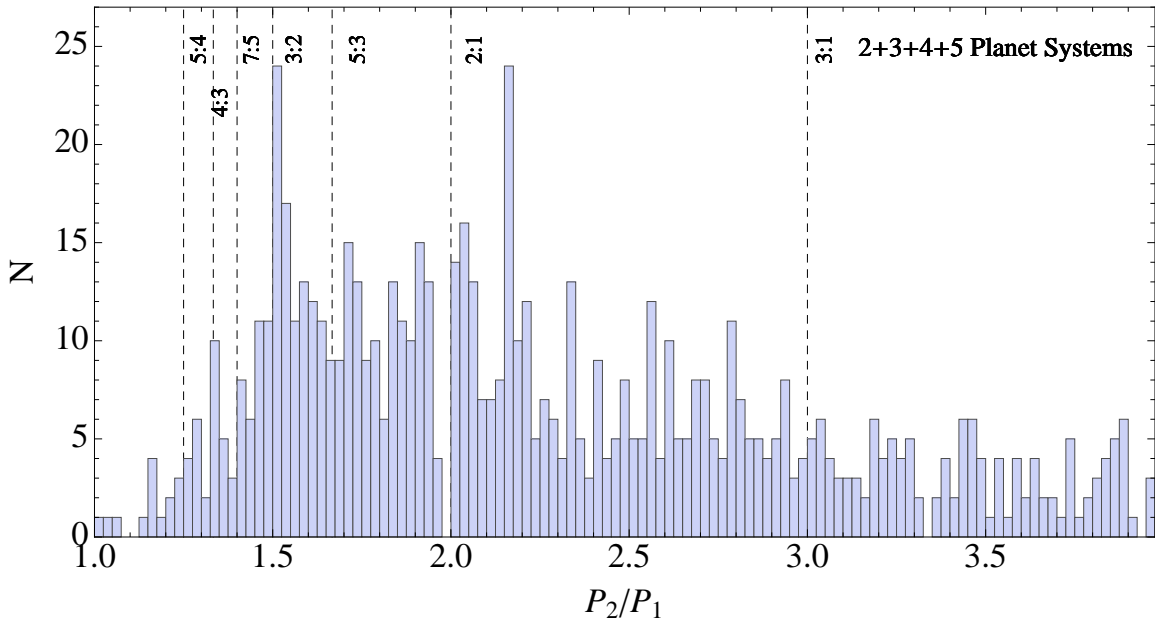, scale=1.00}}
\caption{Histogram showing the period ratios of Kepler planet candidates residing in multiple planet systems as of January 2013. The bin sizes are 0.025. The locations of dominant mean motion resonances are indicated by dashed black lines. Most planets do not reside in or close to resonances. However,
there is a significant excess of planet pairs with period ratios slightly larger than those for exact mean motion resonances. }
\label{fig1}
\end{figure}

As of January 2013, more than 3000 planet candidates have been discovered by the Kepler spacecraft and over 1000 of these reside in systems that contain more than one planet \citep{B13}. Thanks to these remarkable findings, we now have a large statistical sample of multi-planet systems. Their architectures contain important clues concerning planet formation and dynamical evolution. Kepler multi-planet systems display the following characteristics:

$\bullet$ Most planets do not reside in or close to mean motion resonances as shown in Figure \ref{fig1} \citep{FB12}.

$\bullet$ There is a significant excess of planet pairs with period ratios close to but slightly larger (by 1-2\%) than that for exact resonance (see Figure \ref{fig1} and \citet{FB12}).

Several papers have been published offering explainations for the 1-2\% offset from mean motion resonance \citep[e.g.][]{LW12,BM13,PMT13,BP13}. Our investigation has a different focus.  We aim to explain why mean motion resonances are rare.  This is surprising 
because planet-disk interactions leading to orbit migration are expected to result in efficient resonance capture (see section 2.2 and Figure \ref{fig3}).  However, we show that for most Kepler planet pairs, resonance capture is only temporary.  The reason is that 
librations about exact resonance are overstable and lead to passage through resonance on the eccentricity damping timescale which is about two orders of magnitude shorter than that for semi-major axis migration \citep[e.g][]{GT80,Ward88}.

This paper is structured as follows. \S 2 focuses on the planar, circular, restricted three-body problem in the vicinity of a mean motion resonance including resonance capture in the presence of semi-major axis migration and eccentricity damping and the evolution within and escape from resonance.  We show that together, eccentricity damping and orbit migration give rise to an equilibrium eccentricity and that librations around equilibrium are overstable for the majority of Kepler planet masses and lead to passage through resonance. We extend our model to the planar, three-body problem in \S 3.  Results derived from our model are compared with properties of multi-planet systems discovered by the Kepler Spacecraft in \S 4. In \S 5 we discuss reasons for the departure from exact resonance and show that our resonance model yields an excess of pairs with slightly greater than exact resonant period ratios, but that the magnitude of the offset from exact resonance is on average too small to match observations. We compare our findings with previous works that aim to explain the period offset.  Our main conclusions are summarized in \S 6

\section{Resonance in the Circular Restricted Three-Body Problem with Dissipation}
Consider a system of two planets in orbit around a host star. Assume that the outer planet moves on a fixed circular orbit. Close to a first order $j+1:j$ mean motion resonance, the dominant term in the inner planet's disturbing function has resonant argument $\phi$ such that
\begin{equation}\label{e0}
 \phi=(j+1)\lambda'-j\lambda-\varpi, \quad  \ \dot\phi=(j+1)n'-jn-\dot\varpi,\ \quad \ddot\phi=-j\dot n-\ddot\varpi,
\end{equation}
where primes distinguish the outer planet and $\lambda$, $n$, and $\varpi$ denote mean longitude, mean motion and longitude of pericenter. At conjunction, $\lambda' =\lambda$ so $\phi$ is a measure of the displacement of the longitude of conjunction from the inner planet's pericenter.

Lagrange's equations of motion to first order in eccentricity for the inner planet in the vicinity of the resonance read
\begin{equation}\label{e6}
\dot n=3j\beta \mu' en^2\sin\phi-\frac{n}{\tau_{n}}+p\frac{e^2n}{\tau_{e}}\, ,
\end{equation}
\begin{equation}\label{e7}
\dot e=\beta\mu' n\sin\phi-\frac{e}{\tau_{e}}\, ,
\end{equation}
\begin{equation}\label{e8}
\dot\varpi=-\frac{\beta\mu'}{e}n\cos \phi\, ,
\end{equation}
\begin{equation}\label{e9}
\ddot\phi=-3 j^2 \beta \mu'en^2\sin \phi-\left(\frac{\beta \mu' n}{e}\right)^2 \cos\phi \sin \phi-\frac{\beta \mu' n}{e}\sin\phi \dot\phi+\frac{\beta\mu' n}{e\tau_e}\cos\phi+j\frac{n}{\tau_n}-3j\frac{e^2n}{\tau_e}\, ,
\end{equation}
where $e$ and $a$ are eccentricity and semi-major axis, $\mu'$ is the ratio of the outer planet's mass to that of the star, and $\beta\approx 0.8j$.  Although we first examine resonance dynamics without dissipation, terms describing tidally induced migration and  eccentricity damping are included in order to avoid having to repeat these equations later on.   
The final term in equation (\ref{e6}) accounts for the contribution of eccentricity damping to changes in mean motion.  For the particular case of eccentricity damping arising from energy dissipation at constant angular momentum, as applies to tides raised in a planet by its parent star, $p\approx 3$.  For simplicity, we assume $p=3$ throughout the body of our paper but provide results applicable for general $p>0$ in the Appendix.  

For the remainder of this subsection, we neglect both migration and eccentricity damping and work in the limit $\tau_n\, \&\, \tau_e\to \infty$.  
Then equations (\ref{e6})-(\ref{e8}) admit two integrals\footnote{In these integrals, $n$ is evaluated at exact resonance except where it
appears as part of $\dot\phi$.}
\begin{equation}
k(\phi,e^2)= \left(\frac{3}{2}j^2 e^2-\frac{\beta\mu'}{e}\cos \phi \right) +\frac{\dot \phi}{n}\, ,
\end{equation}
and 
\begin{equation}\label{e45}
\mathcal{H}(\phi,e^2)=\left(k e^2-\frac{3}{4}j^2e^4 +2\beta\mu'e\cos \phi \right).
\end{equation}
The integral $k$ exists because with only one resonant argument, variations of $n$ and $e$ are related.  $\mathcal{H}$, the Jacobi constant, is also the Hamiltonian with canonically conjugate momentum, $e^2$, and coordinate, $\phi$.  

\begin{figure} [htp]
\centerline{\epsfig{file=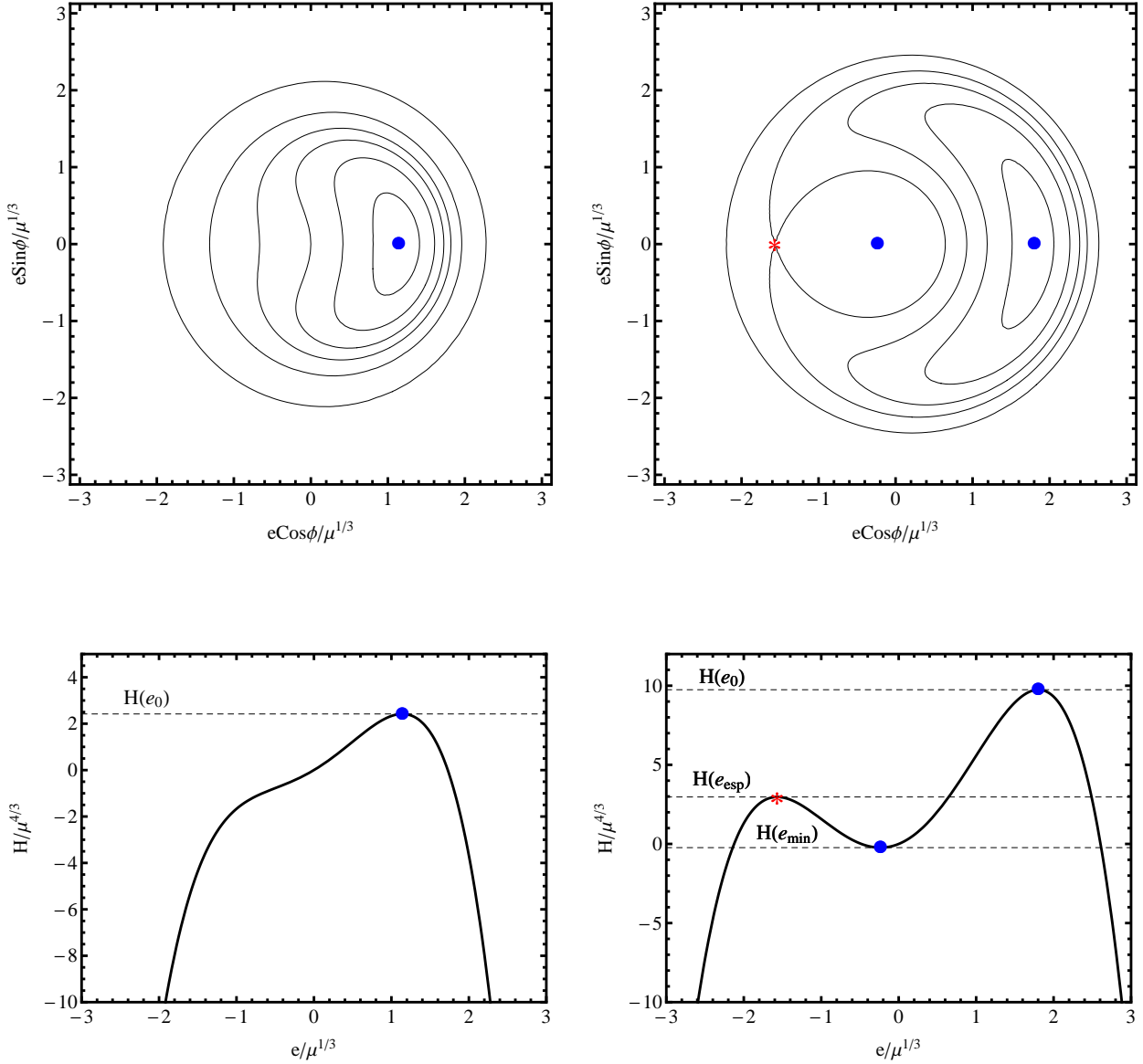, scale=0.75}}
\caption{Contour plots of the Hamiltonian (top) and the corresponding cross section along the $\sin\phi=0$ axis (bottom) for $k=0.5 k_{crit}$ (left) and $k=2 k_{crit}$ (right). Negative values on the bottom x-axes correspond to $\phi=\pi$. Contours on the left are shown for $H(e,\phi)/\mu^{4/3} \simeq 2, 1, 0, -1, -3, -10$. Those on the right correspond to $H(e,\phi)/\mu^{4/3} \simeq 9, 7, 5, -1$. Stable fixed points are marked by blue circles and the unstable one by a red asterisk.}
\label{fig11}
\end{figure}

The topology of the phase-space as defined by $\mathcal{H}$ for fixed $k$ changes abruptly across $k=k_{crit}$, where
\begin{equation}
k _{crit}= \frac{3^{4/3}}{2}\left(j\beta \mu'\right)^{2/3}\sim j^{4/3}\mu'^{2/3}\, .
\end{equation}
For $k<k_{crit}$, there is one (stable) fixed point whereas there are three (two stable and one unstable) fixed points for $k>k_{crit}$.\footnote{The fixed points correspond to the real roots of the expression for $k=0$ with $\phi=0$.}  These 
distinct topologies are illustrated in Figure \ref{fig11}.  For $k>k_{crit}$ the level curve emanating from the unstable fixed point is appropriately called a separatrix because it separates the regions surrounding the two stable fixed points. The stable fixed point at
\begin{equation}
\phi=0 \quad {\rm and}\quad e_0=\frac{2}{j^{2/3}}
\left(\frac{\beta\mu'}{3}\right)^{1/3}\left(k\over k_{crit}\right)^{1/2}=\frac{\beta\mu' n}{jn-(j+1)n'}
\end{equation}
is present for all $k$ and corresponds to a periodic orbit with $jn>(j+1)n'$.  It  is located at the global maximum of $\mathcal{H}$ and owes its stability to the Coriolis acceleration in the frame rotating with angular velocity $n'$.  The other stable fixed point at
\begin{equation}
\phi=\pi \quad {\rm and}\quad e_{min}=\frac{1}{j^{2/3}}\left(\frac{\beta\mu'}{3}\right)^{1/3}\left(k\over k_{crit}\right)^{1/2}\left(\cos\theta-3^{1/2}\sin\theta\right)=\frac{\beta\mu' n}{(j+1)n'-jn}\, ,
\end{equation}
with
\begin{equation}
\theta=\frac{1}{3}\tan^{-1}[(k/k_{crit})^3-1]
\end{equation}
only appears for $k>k_{crit}$ and describes a periodic orbit with $jn<(j+1)n'$. It sits at a local minimum of $\mathcal{H}$.  Finally, the unstable fixed point at 
\begin{equation}
\phi=\pi \quad {\rm and}\quad e=\frac{1}{j^{2/3}}\left(\frac{\beta\mu'}{3}\right)^{1/3}\left(k\over k_{crit}\right)^{1/2}\left(\cos\theta+3^{1/2}\sin\theta\right)=\frac{\beta\mu'}{(j+1)n'-jn}
\end{equation}
lies on a saddle point. The stable and unstable fixed points at $\phi=\pi$ bifurcate at $k=k_{crit}$ where $e_{crit}=(\beta\mu'/3 j^2)^{1/3}$.  These values correspond to the boundary between stable and unstable linear oscillations around $\phi=\pi,\,e=\beta\mu' n/((j+1)n'-jn)$ as can be verified from equation (\ref{e9}).  In anticipation of later discussion, we make note of the asymmetry of the stable fixed points on opposite sides of exact resonance. 

Note that near resonance, $e=O(\mu'^{1/3})$ and $d/dt=O(\mu'^{2/3}n)$.  Thus the first and second terms on the rhs of equation (\ref{e9}) dominate over the third for small amplitude librations (i.e., $\phi = \delta \phi << 1$).   
Moreover,  librations of $\phi$ are centered on $\phi=0,\, e=e_0$ \citep[e.g.][]{MD99} where
\begin{equation}\label{e104}
e_0=\frac{\beta\mu' n}{jn-(j+1)n'}\, .
\end{equation}
From equation (\ref{e9}), it follows that the frequency of small librations, $\omega$, satisfies
\begin{equation}\label{e125}
\frac{\omega^2}{n^2}=3j^2\beta\mu' e_0+\left(\beta\mu'\over e_0\right)^2\, .
\end{equation}

\subsection{Planet-Disk Interactions}\label{PDI}
Up to this point, migration and eccentricity damping were merely parametrized by their respective timescales $\tau_n$ and $\tau_e$.  For dissipation due a planet's interaction with a protoplanetary disk, $\tau_{n}$ and $\tau_{e}$ are given by
\begin{equation}\label{e17}
\frac{1}{\tau_{n}} \equiv \frac{1}{n}\frac{dn}{dt} \sim \mu\mu_d \left(\frac{a}{h}\right)^2n\, ,
\end{equation}
\begin{equation}\label{e18}
\frac{1}{\tau_{e}} \equiv \frac{1}{e} \frac{de}{dt} \sim \mu \mu_d \left(\frac{a}{h}\right)^4n\, ,
\end{equation}
where $\mu_d=\Sigma_d a^2/M_*$ is the disk to star mass ratio and $h$ is the disk's scale height.
Here we have assumed that the planet's mass is too small for it to clear a gap in the disk. 

The migration rate is based on the balance of torques at principal Lindblad resonances located interior and exterior to the planet \citep{GT80, Ward86}.  Eccentricity damping in a gapless disk is dominated by first order, coorbital Lindblad resonances \citep{Ward88, Artymowicz93}.  These come in pairs with individual members approximately equally spaced interior and exterior to the coorbital location. Consequently, each pair damps the planet's orbital energy at nearly constant angular momentum.  However, remote first order Lindblad resonances along with coronation resonances also contribute to
eccentricity excitation and damping.  Thus the coefficient $p$ in equation (\ref{e6}) may deviate from $3$.
For example, the plant-disk model investigated by \citet{TW04} yields $p \sim 1.3$. Overstable librations require that the final term in equation (\ref{e6}) be positive as shown by equation (\ref{e713}).

\subsection{Capture Into Resonance}

\begin{figure} [htp]
\centerline{\epsfig{file=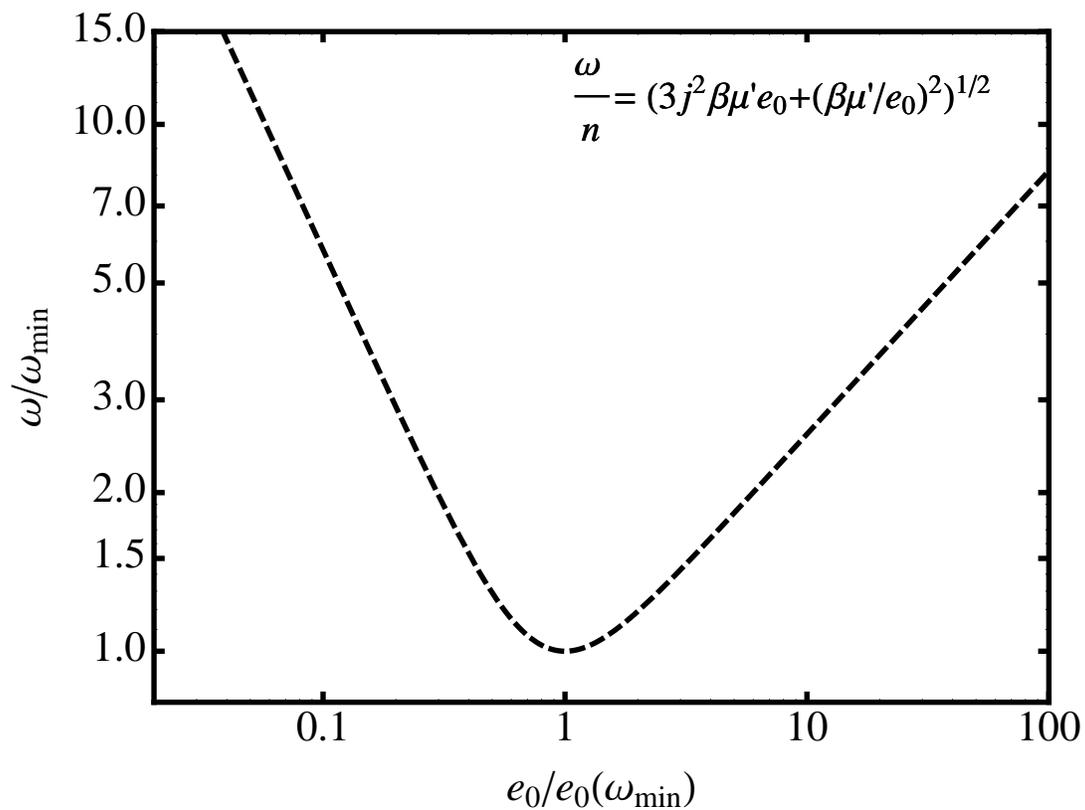, scale=1.4}}
\caption{Frequency of small amplitude libration about the stable fixed point as a function of eccentricity, $e_0$. Capture into resonance requires that the migration timescale across the resonance be longer than the libration timescale. This condition is given by $\dot n/\Delta n \lesssim \omega_{min}$, where $\Delta n$ is the resonance width.}
\label{fig2}
\end{figure}

Next we consider conditions under which the inner planet would be captured into the 2:1 resonance if it were migrating outward at a rate ${|\dot n|}/n=1/\tau_n$.  Capture would be guaranteed provided the orbital eccentricity of the inner body far from the resonance were sufficiently small and the rate of outward migration sufficiently slow.  Let us assume that
the former condition is satisfied.  Then upon approach to resonance, the inner planet would stay close to the stable fixed point at $\phi=0$ and $e_0=\beta\mu' n/(jn-(j+1)n')$.  As the resonance is approached, $\omega$ first decreases and then increases (see Fig. \ref{fig2}).  Capture is most problematic during passage through 
\begin{equation}\label{e100}
\frac{\omega_{min}}{n}=\frac{3^{4/3}}{2^{1/3}}\left(j\beta\mu'\right)^{2/3}\sim \left(j^2\mu'\right)^{2/3}\, ,
\end{equation}
at which point
\begin{equation}\label{e101}
e_0(\omega_{min})=\left(2\beta\mu'\over 3j^2\right)^{1/3}\sim \left(\mu'\over j\right)^{1/3}\, ,
\end{equation}
and
\begin{equation}\label{e102}
\frac{\Delta n}{n}\equiv 1-\frac{(j+1)n'}{jn}=\left(3\over 2j\right)^{1/3}\left(\beta\mu'\right)^{2/3}\sim j^{1/3}\mu'^{2/3}\, .
\end{equation}
To cross the resonance width, $\Delta n/n\sim  j^{1/3}\mu'^{2/3}$ given by equation (\ref{e102}) while migrating at rate $|\dot n|/n=1/\tau_n$ takes $\Delta t\sim j^{1/3}\mu'^{2/3}\tau_n$.  Capture would be assured provided $\omega_{min}\Delta t\sim j^{5/3}\mu'^{4/3}n\tau_n\gg 1$.  Careful analytical and numerical calculations sharpen this result to
\begin{equation}\label{e103}
j^{5/3}\mu'^{4/3}n\tau_n\geq 2.5\, .
\end{equation}
For smaller $\tau_n$,  temporary capture could occur at larger separations from resonance where $\omega>\omega_{min}$ but the resonance lock would be broken before $\omega_{min}$ were reached.  With $\tau_n$ given by equation (\ref{e17}), the condition of certain capture becomes
\begin{equation}\label{e3}
\mu' \gtrsim \frac{\mu_d^3}{j^5}\left(\frac{a}{h}\right)^6.
\end{equation}

Having defined sufficiently slow migration, we do the same for sufficiently small initial eccentricity.  The key is to make use of the adiabatic invariant 
\begin{equation}\label{e55}
AI=\oint_C \, d\phi\,e^2\, .
\end{equation}
For motion on the separatrix at $k=k_{crit}$, it can be shown that $AI_{sep}=12\pi (\beta\mu'/3j^2)^{2/3}$.  Thus if well before a planet encounters the resonance, its orbital eccentricity were smaller than
\begin{equation}
e_{cap}=\left(AI_{sep}\over 2\pi\right)^{1/2}=1.7\left(\beta\mu' \over j^2\right)^{1/3}\sim \left(\mu' \over j\right)^{1/3}\, ,
\end{equation}
then at $k=k_{crit}$, $\mathcal H$ would be larger than ${\mathcal H}_{sep}$ and capture would be guaranteed.

\begin{figure} [htp]
\centerline{\epsfig{file=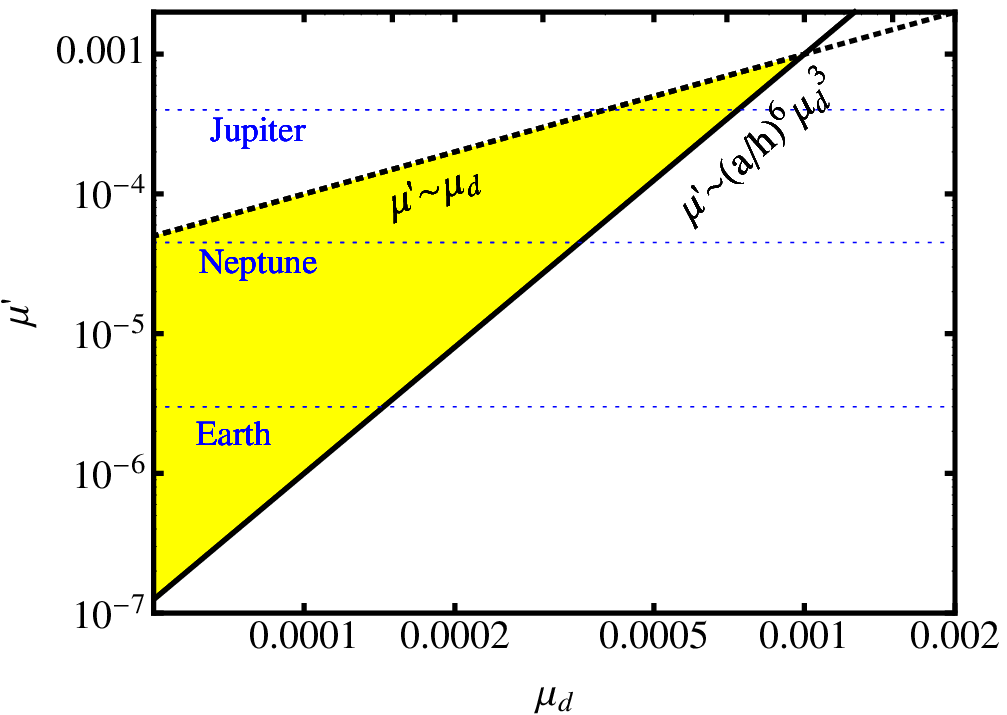, scale=1.4}}
\caption{Criteria for capture into the 2:1 resonance for planets migrating in a protoplanetary disk as a function of disk to star mass ratio, $\mu_d \sim \Sigma_d a^2/M_*$, and planet to star mass ratio, $\mu'$. The solid black line delineates the lower boundary above which capture can occur as expressed by equation (\ref{e3}). The dotted black line corresponds to equality of planet and local disk mass. The effect of the disk on planet migration diminishes if the local disk mass is less than the planet's mass. The yellow shaded area marks the parameter space within which capture into resonance can take place. As indicated by equation (\ref{e3}), the minimum value of $\mu'$ required for capture into higher $j$ resonances is lower than shown here.}\label{fig3}
\end{figure}

Conditions for resonance capture appropriate to planets migrating in a protoplanetary disk are displayed in Figure \ref{fig3} as a function of the local disk and planet masses. The local disk mass, $\mu_d$, is that which resides within a factor of two of the planet's orbital radius.  For a total disk mass equal to $10^{-2}M_*$ distributed with a radial surface density profile $\propto a^{-1}$, the local disk mass at 0.5 AU is two orders of magnitude smaller than the local disk mass at 50 AU. Figure \ref{fig3} indicates that resonance capture will only occur for local disk masses less than $10^{-3}M_*$.  Thus in a typical disk with total mass $10^{-2}M_*$ and an $a^{-1}$ surface density profile, only planets within about 10AU would be caught into resonance. Figure \ref{fig3} further suggests planets with masses between those of Earth and Neptune which orbit within 1AU have the best chances to capture smaller planets.   

To a significant extent, the current situation regarding mean motion resonances among pairs of exoplanets is analogous to that which pertained to mean motion resonances in the satellite systems of Jupiter and Saturn more than half a century ago. 
A early paper by \cite{RO54} called attention to the overabundance of orbital resonances among the satellite systems of Jupiter and Saturn.  \cite{Goldreich65} showed
that slow convergent migration driven by tides raised in a planet would lead to the formation of mean motion resonances stabilized by the transfer of angular momentum from the inner to the outer satellite. {Early work emphasized slow migration because it is appropriate to cases in which tidal torques drive the expansion of satellite orbits.} The probability of capture into resonance for slow convergent migration was first solved by \cite{Yoder79}.  Subsequently, simpler derivations emphasizing the role of adiabatic invariance were provided in \cite{Henrard82, BG84}.  Recent investigations focused on the maximum migration rate permitting capture.  The analytic derivation by \cite{Friedland01} is in close agreement with results from numerical experiments \citep{Quillen06, OK13}.  

\subsection{Evolution in Resonance in the Presence of Migration}
Provided that the migration is sufficiently slow such that resonant capture occurs, the orbital eccentricity $e_0$ keeps growing on a timescale comparable to the migration timescale, $\tau_n$, as the planet moves deeper into resonance \citep[e.g.][]{M93,LP02}. Once the eccentricity has grown to order unity, the system becomes unstable leading to passes through resonance, collision or ejection \citep[e.g.][]{LP02}. Since the eccentricity growth in resonance occurs on a timescale comparable to $\tau_n$ this implies that planets should share their time roughly equally between residing in mean motion resonances and migrating between them. Thus at any given time, about half of all planet pairs should occupy mean motion. This implies that in the presence of migration that about half of all planet pairs should occupy mean motion resonances at any given moment in time, which is not consist with the observed period rations of Kepler planet pairs (see Figure \ref{fig1}). As we show below, eccentricity damping provides a simple solution to this conundrum.

\subsection{Effects of Eccentricity Damping}\label{ssedamp}
Evolution in resonance is very different in the presence of eccentricity damping. This is because as the planet evolves deeper into resonance the associated increase in its eccentricity can be balanced by a decrease due to eccentricity damping. Equations (\ref{e6}) and (\ref{e7}) imply the existence of an equilibrium eccentricity and resonance phase angle given by
\begin{equation}\label{e106}
e_{eq}= \left( \frac{\tau_{e}}{3(j+1)\tau_{n}}\right)^{1/2} 
\end{equation} 
and 
\begin{equation}\label{e107}
\sin\phi_{eq} = \frac{e_{eq}}{\beta\mu'\tau_{e} n}.
\end{equation}
No such equilibrium exists in the presence of migration alone.

In what follows we show that evolution in resonance with eccentricity damping and migration has three possible outcomes which depend upon the values of $\beta\mu'$ and $\tau_e/\tau_n$. 

\begin{figure}[htp]
\centering
\begin{minipage}[b]{0.45\linewidth}
\includegraphics[width=7.5cm,height=16cm]{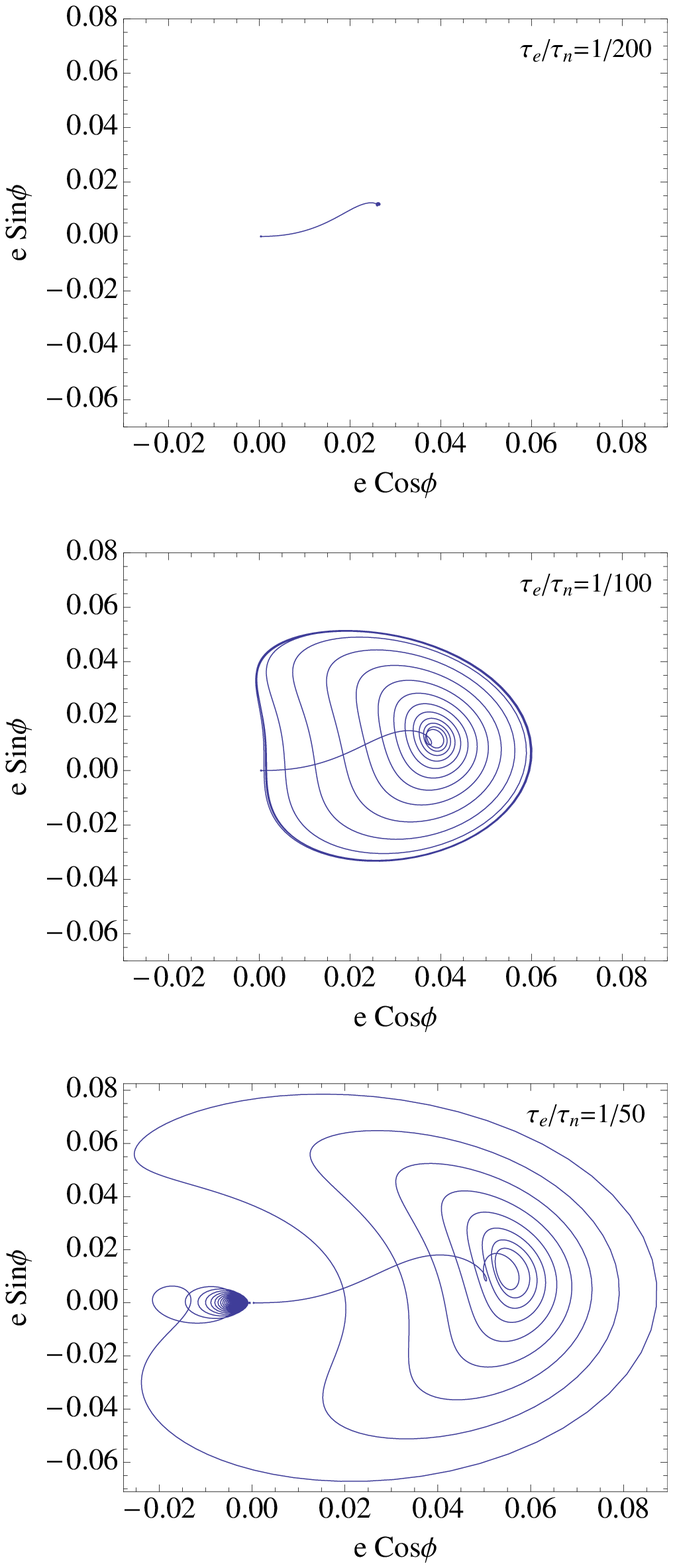}
\end{minipage}
\quad
\begin{minipage}[b]{0.45\linewidth}
\includegraphics[width=7.5cm,height=16cm]{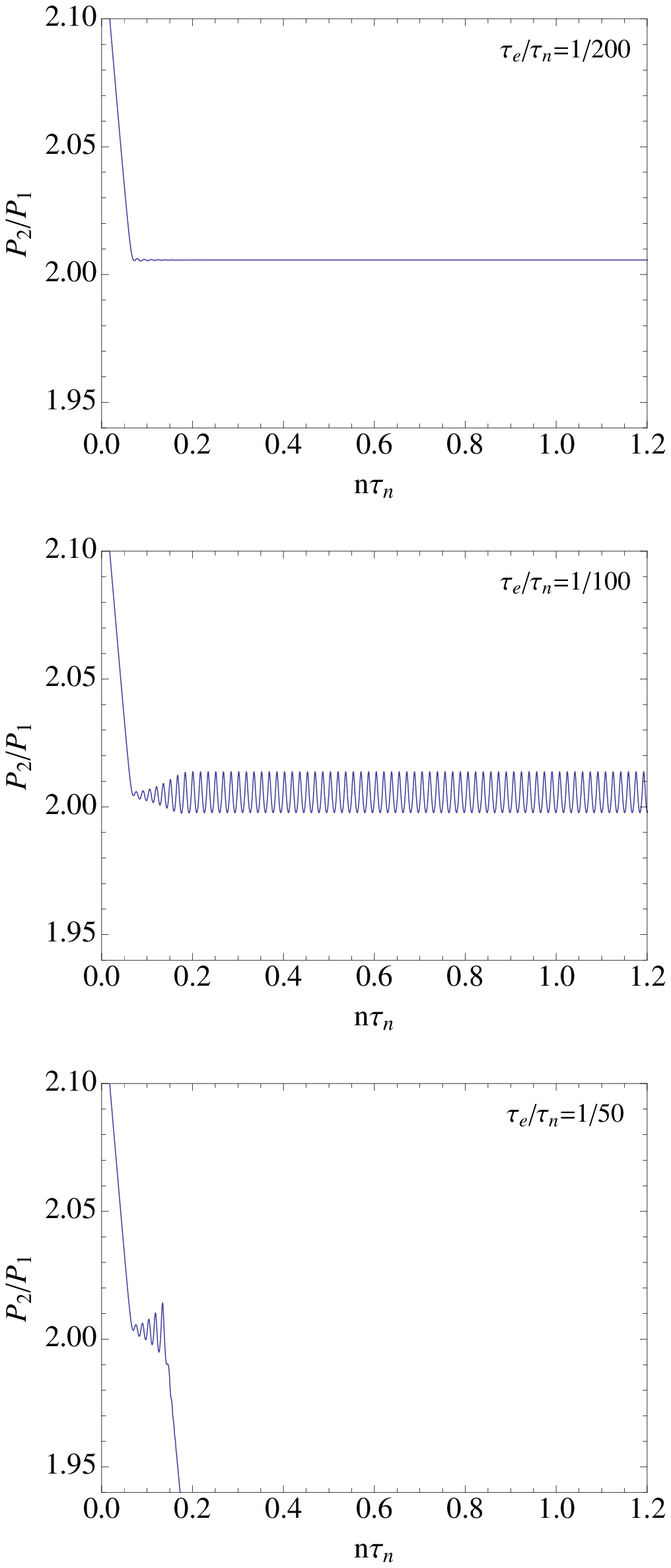}
\end{minipage}
\caption{Examples of evolution in resonance for $\mu'=10^{-4}$, $n\tau_{n}=10^5$, and different values of $\tau_{e}/\tau_{n}$. Left panel: Librations are centered on $\phi_{eq}\simeq e_{eq}/(\beta\mu'\tau_{e} {n})$ and the offset of $\phi$ from 0 is clearly visible in each plot.  In the upper panel, $\tau_e/\tau_n=0.005$.  The final state corresponds to permanent resonant capture at $\phi_{eq},\, e_{eq}$ (case 1).  In the middle panel, $\tau_e/\tau_n=0.01$.  The libration amplitude grows and then saturates but again, resonant capture is permanent (case 2).  In the bottom panel, $\tau_e/\tau_n=0.02$. The libration amplitude grows and the separatrix is crossed leading to damped inner circulation and escape from resonance (case 3). Right panel: Evolution of the planet pair period ratio as a function of time corresponding to permanent capture with no librations (case 1) top panel, to permanent capture with finite amplitude librations (case 2) middle panel, and to temporary capture and escape (case 3) bottom panel. Note, escape from resonance occurs on a timescale comparable to $\sim \tau_e$ rather than $\tau_n$.
}
\label{fig4}
\end{figure}

1) For 
\begin{equation}\label{e108}
\mu' > \frac{j}{\sqrt{3}(j+1)^{3/2}\beta}\left(\tau_{e}\over \tau_{n}\right)^{3/2}\, ,
\end{equation}
the planet is permanently trapped in resonance at the fixed point $\phi=\phi_{eq}$ and $e=e_{eq}$.  An example is shown in the upper panel of Figure \ref{fig4}.

2) For 
\begin{equation}\label{e109}
\frac{j^2}{8\sqrt{3}(j+1)^{3/2}\beta}\left(\tau_{e}\over \tau_{n}\right)^{3/2} < \mu' < \frac{j}{\sqrt{3}(j+1)^{3/2}\beta}\left(\tau_{e}\over \tau_{n}\right)^{3/2}\, ,
\end{equation}
the planet is permanently caught in resonance and its libration amplitude saturates at a finite value.  The middle panel of Figure \ref{fig4} displays an example.

3) For 
\begin{equation}\label{e110}
\mu' < \frac{j^2}{8\sqrt{3}(j+1)^{3/2}\beta}\left(\tau_{e}\over \tau_{n}\right)^{3/2}\, ,
\end{equation}
the planet is caught in resonance, but escapes on timescale $\tau_e$. An example is provided in the lower panel of Figure \ref{fig4}.

Note that the boundary between permanent capture and temporary capture leading to escape shifts monotonically to smaller $\mu'\propto j^{-1/2}$ with increasing $j$.  Moreover, the intermediate range of $\mu'$ corresponding to finite amplitude librations shrinks with increasing $j$ and does not exist for $j\geq 8$. 

\begin{figure} [htp]
\centerline{\epsfig{file=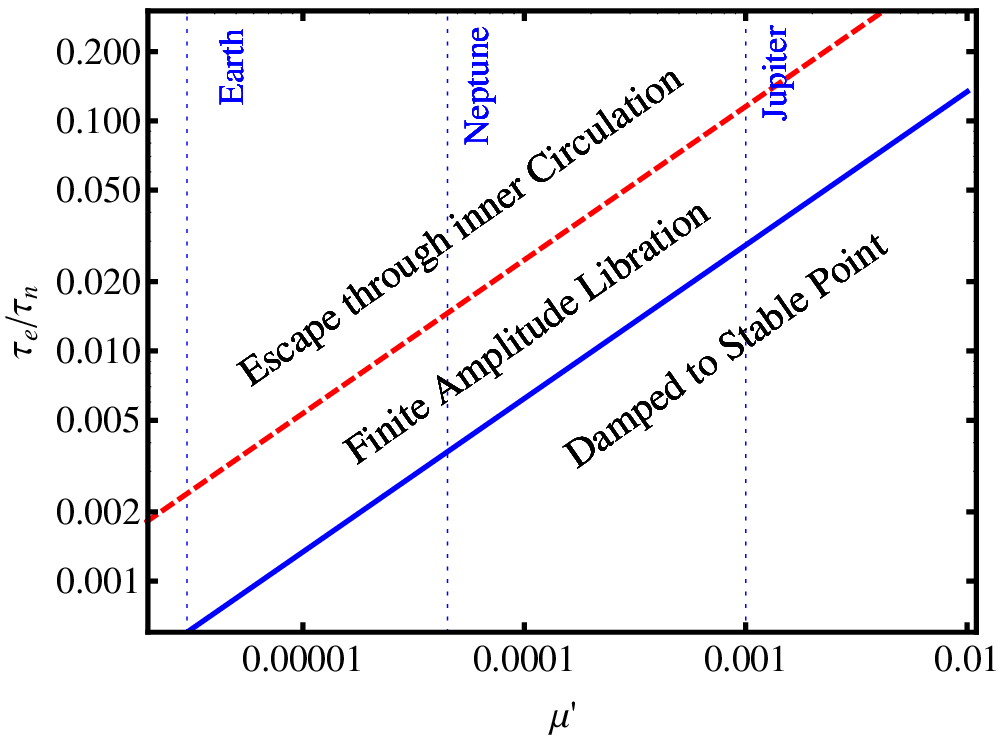, scale=1.4}}
\caption{Different outcomes for the evolution in resonance for various valves of $\mu'$ and $\tau_{e}/\tau_{n}$. The solid blue line marks the transition from libration damping below to libration growth above. The red dashed line corresponds to the transition from finite amplitude librations below to escape from resonance above. Planet pairs with parameters that fall below the dashed red line are permanently captured in resonance. The rest escape on a timescale $\tau_{e}$. The vertical lines show solar system planet to sun mass ratios for Earth, Neptune and Jupiter. Note that for a given resonant planet pair, $\mu'$ corresponds to the larger of the two masses.}
\label{fig5}
\end{figure}

Figure \ref{fig5} summarizes the three different outcomes for the evolution in resonance for various planet to star mass ratios and as a function of $\tau_{e}/\tau_{n}$ for a 2:1 mean motion resonance.

Below we derive analytic results for different evolutions in resonance.

\subsubsection{Instability: Transition from Damping to Growth}\label{ssoverstab}

Linearizing equations (\ref{e0}) for $\dot\phi$, (\ref{e6}) for $\dot n$, and (\ref{e7}) for $\dot e$ around the equilibrium given by $\phi_{eq}$ and $e_{eq}$ and setting $d/dt=s$ yields a system of three homogeneous linear equations in the variables $\delta\phi$, $\delta n$ and $\delta e$.  The determinant of the coefficient matrix is a 
cubic polynomial in $s$ with real coefficients whose roots are the eigenvalues.  A standard exercise in
algebra reveals that one root is negative and that the other two are a complex conjugate pair.  For slow migration and eccentricity damping, $\omega\tau_e\gg 1$ and $\omega\tau_n\gg 1$, the imaginary parts of the complex roots are essentially $\pm i\omega$ evaluated at $e_0=e_{eq}$ and the real part is well-approximated by
\begin{equation}\label{e13}
s=\frac{1}{\tau_{e}}\left(\frac{n}{\omega_{eq}}\right)^2\left(3j\beta \mu'e_{eq}-\left(\frac{\beta\mu'}{e_{eq}}\right)^2\right)\, .
\end{equation}
Thus the evolution in resonance switches from damping to growth at $e_{eq}=(\beta\mu'/3j)^{1/3}$. Since $e_{eq} = \left(\tau_e/(3(j+1)\tau_n)\right)^{1/2}$,  the libration amplitude grows if
\begin{equation}\label{e111}
\mu'<\frac{j}{\sqrt{3}(j+1)^{3/2}\beta}\left(\tau_{e}\over \tau_{n}\right)^{3/2}\, .
\end{equation}
Furthermore, the growth timescale from equation (\ref{e13}) is of order $\tau_e$.  Thus, escape from resonance, if it occurs, will happen on a timescale $\tau_e$. 

At a basic level, overstability can arise because $e_0$ corresponds to a maximum of $\mathcal H$ and eccentricity damping causes the system to slide downhill.  It is necessary to dig a little deeper to understand the origin of the transition from damping to overstability. Once again, an adiabatic invariant comes into play, this time involving small librations around $\phi=0,\,e=e_0$.\footnote{This adiabatic invariant only applies for slow migration.  It is broken by eccentricity damping.}  A simple calculation using equation (\ref{e55}) yields
\begin{equation}
AI_{e_0}=2\pi\beta\mu'\frac{n}{\omega_0}e_0\Phi^2\, ,
\end{equation}
where $\Phi$ is the amplitude of the libration and $\omega_0$ its frequency. For $e_0\ll \mu'^{1/3}$, $\omega_0/n\sim \mu'/e_0$ whereas for $e_0\gg \mu'^{1/3}$, $\omega_0/n\sim (\mu' e_0)^{1/2}$. By itself, convergent migration results in a monotonic increase of $e_0$. The adiabatic invariant then implies
that the libration amplitude decreases as $e_0$ increases.  But for $e_0\ll \mu'^{1/3}$, $\Phi\propto e_0^{-1}$ whereas for $e_0\gg \mu'^{1/3}$, $\Phi\propto e_0^{-1/4}$.  Eccentricity damping leads to an increase in $\Phi$ which migration can overcome for $e_0\ll \mu'^{1/3}$ but not for $e_0\gg \mu'^{1/3}$.

\subsubsection{Escape from Resonance}
Following capture in resonance, escape involves damping to the stable fixed point  at $\phi=\pi,\, e=e_{min}$.  This can only occur if $k>k_{crit}$ and involves crossing the inner branch of the separatrix which leads to damped circulation about the enclosed fixed point.  As shown at the beginning of \S 2, the separatrix exists for $e_0 \geq 2(\beta\mu'/3j^2)^{1/3}$. Equating $e_{eq}=(\tau_{e}/(3(j+1)\tau_{n}))^{1/2}$ to $e_0$ demonstrates that resonance escape can occur provided
\begin{equation}\label{112}
\mu' < \frac{j^2}{8\sqrt{3}(j+1)^{3/2}\beta}\left(\tau_{e}\over \tau_{n}\right)^{3/2}\, .
\end{equation}

\begin{figure} [htp]
\centerline{\epsfig{file=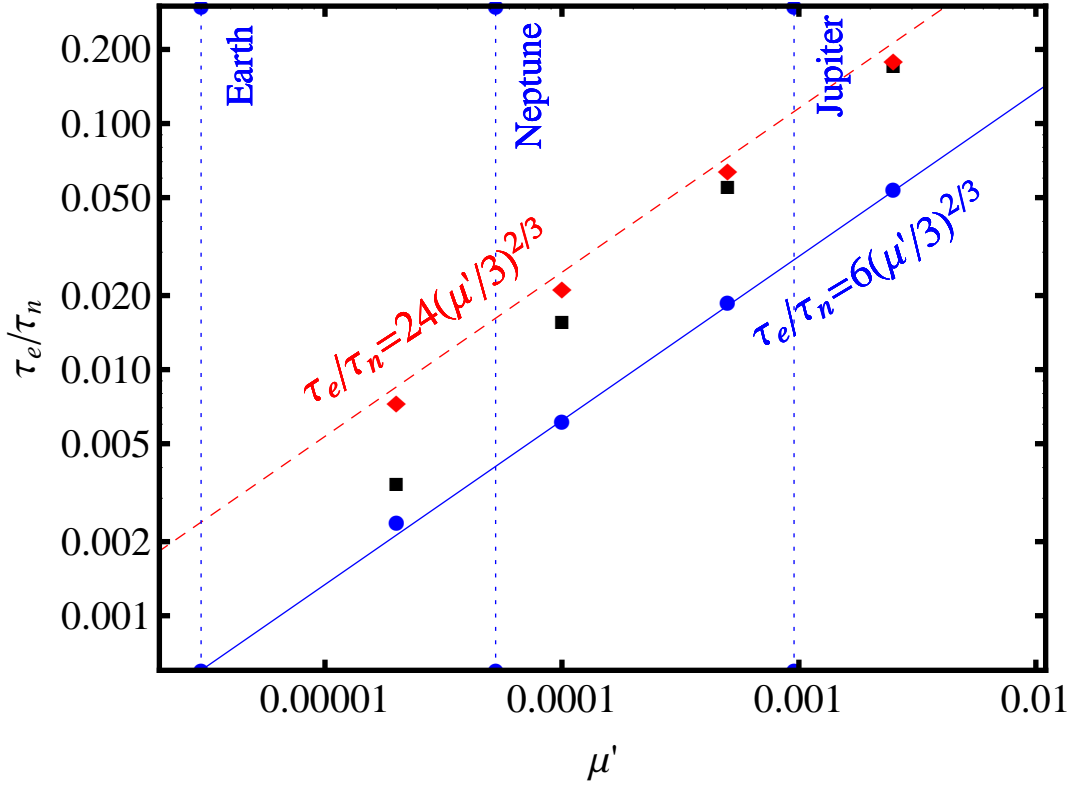, scale=1.4}}
\caption{Outcomes for evolution in the $2:1$ resonance as a function of $\mu'$ and strength of dissipation. The solid blue line corresponds to transition from librations damping (below) to growing (above)  as calculated analytically. The red dashed line corresponds to transition from finite amplitude librations (below) to escape from resonance (above) calculated analytically in the limit that the damping timescale is long compared to the libration timescale. The blue points represent numerical results for which the transition from damping to growth occurred.  Values  obtained for both $\tau_{n} n= 10^5$ and $\tau_{n} n=10^6$ are indistinguishable.  Numerical results for the transition from finite amplitude growth to escape from resonance are marked by black squares for $\tau_{n} n= 10^5$ and red diamonds for $\tau_{n} n=10^6$.  Although the transition from damping to growth seems insensitive to the strength of the dissipation, the transition between finite amplitude libration and escape from resonance systemically
departs from the analytic prediction with decreasing $\mu'$ for $\tau_n n=10^5$.}
\label{fig7}
\end{figure}

Different outcomes for evolution in resonance as a function of $\beta\mu'$ and $\tau_e/\tau_n$ are illustrated in Figure \ref{fig7} for the particular case of a $2:1$ resonance.  Plotted points show results from direct numerical integrations of the resonant equations of motion and the solid and dashed lines present our analytic results. The blue circles, which mark the transition from damping to growth, are indistinguishable for $\tau_{n} n= 10^5$ and $\tau_{n} n=10^6$ and are in accord with our analytic limit (solid blue line). The black squares and red diamonds show the transition from finite amplitude growth to escape from resonance for $\tau_{n} n= 10^5$ and $\tau_{n} n=10^6$, respectively. Although the numerical results for $\tau_{n} n=10^6$ are in close agreement  with our analytic expression (dashed red line), those for $\tau_{n} n=10^5$ fall well below the analytic limit for $\mu'\lesssim 10^{-4}$. The latter is not surprising.  These combinations of $\tau_n$ and $\mu'$ do not conform to the basic assumption of our analytic treatment, namely that damping only makes a small perturbation to the dissipation free motion. In particular, we suspect that the problem in this case is that the eccentricity damping timescale is comparable to the libration period.

\section{The Planar Three-Body Problem: A More Realistic Example}

So far we have assumed that the more massive outer planet moves on a fixed circular orbit and that $m_1/m_2 << 1$.\footnote{Subscript '1' and '2' denote quantities associated with the inner planet and outer planet, respectively.}  In this case there is only one resonant argument.  Here we investigate a scenario in which both the inner and outer planet are in resonance. We still assume that the outer planet with mass $m_2$ is the more massive one thus ensuring that type-I migration is convergent (eqn. \ref{e17}). 

\subsection{Two Planet System with Convergent Migration and Eccentricity Damping}
Consider two planets that orbit their host star in the vicinity of a 2:1 mean motion resonance. The first order resonant terms in the disturbing function for a 2:1 mean motion resonance for the inner and outer planets are, respectively,
\begin{equation}
\mathcal{R}_1=\alpha a_1^2\mu_2 n_1^2\left(-be_1\cos \phi_1+ce_2 \cos \phi_2\right)
\end{equation}
and
\begin{equation}
\mathcal{R}_2=a_2^2\mu_1 n_2^2\left(-be_1\cos \phi_1+ce_2 \cos \phi_2\right)\, ,
\end{equation}
where $\alpha=a_1/a_2=0.630$, $b=1.190$, and $c=0.428$ and $\phi_1=2\lambda_2-\lambda_1-\varpi_1$ and $\phi_2=2\lambda_2-\lambda_1-\varpi_2$. In the presence of migration and eccentricity damping
\begin{equation}\label{e200}
\dot n_1=-\frac{3}{a_1^2}\frac{\partial \mathcal{R}_1}{\partial\lambda_1} +\frac{n_1}{\tau_{n1}}+\frac{3n_1e_1^2}{\tau_{e1}} =3\alpha \mu_2n_1^2\left(be_1\sin\phi_1-ce_2\sin\phi_2\right)+\frac{n_1}{\tau_{n1}}+\frac{3n_1e_1^2}{\tau_{e1}}\, ,
\end{equation}
\begin{equation}
\dot n_2=-\frac{3}{a_2^2}\frac{\partial \mathcal{R}_2}{\partial\lambda_2} +\frac{n_2}{\tau_{n2}}+\frac{3n_2e_2^2}{\tau_{e2}}=-6 \mu_1n_2^2\left(be_1\sin\phi_1-ce_2\sin\phi_2\right)+\frac{n_2}{\tau_{n2}}+\frac{3n_2e_2^2}{\tau_{e2}}\, ,
\end{equation}
\begin{equation}\label{e46}
\dot e_1=-\frac{1}{n_1a_1^2e_1}\frac{\partial \mathcal{R}_1}{\partial\varpi_1}-\frac{e_1}{\tau_{e1}}=\alpha b\mu_2 n_1\sin\phi_1-\frac{e_1}{\tau_{e1}}\, ,
\end{equation}
\begin{equation}
\dot e_2=-\frac{1}{n_2a_2^2e_2}\frac{\partial \mathcal{R}_2}{\partial\varpi_2}-\frac{e_2}{\tau_{e2}}=-c\mu_1 n_2\sin\phi_2-\frac{e_2}{\tau_{e2}}\, ,
\end{equation}
\begin{equation}\label{e47}
\dot \varpi_1=-\frac{1}{n_1a_1^2e_1}\frac{\partial e_1}{\partial\varpi_1}=- \alpha b\mu_2 n_1\frac{\cos\phi_1}{e_1}\, ,
\end{equation}
\begin{equation}
\dot \varpi_2=-\frac{1}{n_2a_2^2e_2}\frac{\partial e_2}{\partial\varpi_1}=c\mu_1 n_2\frac{\cos\phi_2}{e_2}\, .
\end{equation}
Since we are dealing with a 2:1 mean motion resonance,  $\phi_1=2\lambda_2-\lambda_1-\varpi_1$ and $\phi_2=2\lambda_2-\lambda_1-\varpi_2$, which simplify to $\phi_1=\sigma-\varpi_1$ and $\phi_2=\sigma-\varpi_2$ where
\begin{equation}\label{e201}
\dot \sigma = 2n_2 -n_1\, .
\end{equation}
After substituting for $\phi_1$ and $\phi_2$, equations (\ref{e200}) to (\ref{e201}) yield a set of 7 first order differential equations in the 7 independent variables $\sigma, n_1, n_2, e_1, e_2, \varpi_1, \varpi_2$.

\begin{figure} [htp]
\centerline{\epsfig{file=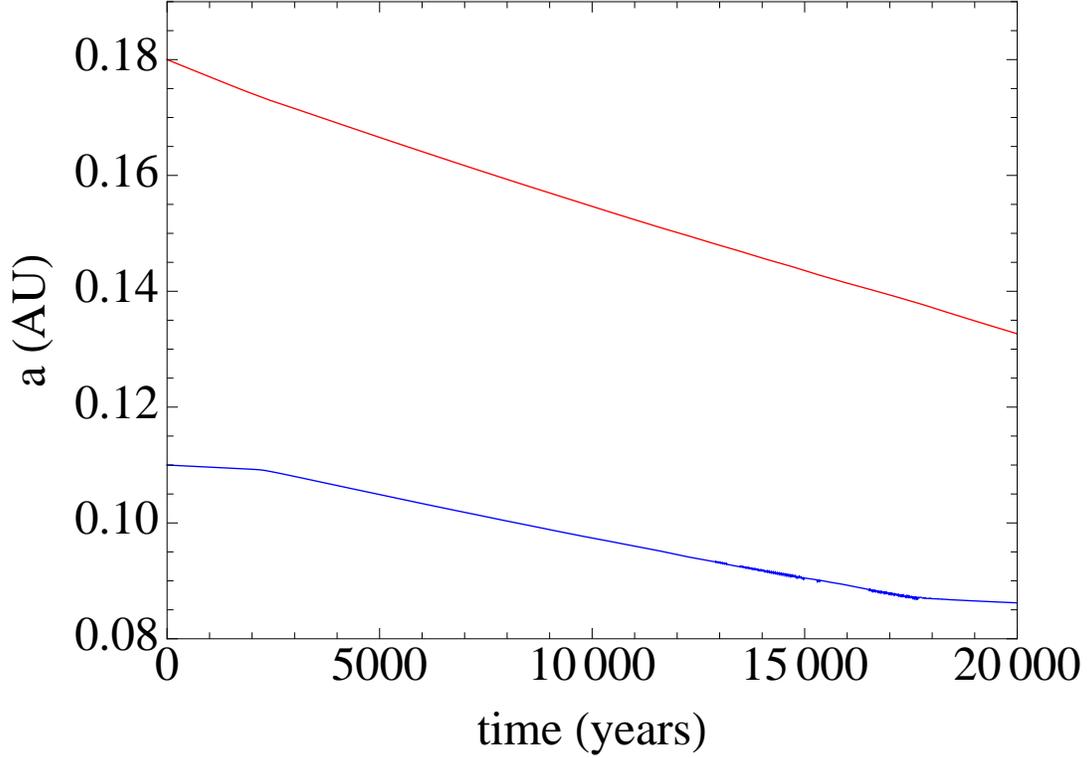, scale=1.4}}
\caption{Semi-major axis as a function of time for planet 1 (blue) and planet 2 (red). Convergent migration leads to temporary capture into 2:1 mean motion resonance, but the resonance is broken after a few eccentricity damping timescales, $\tau_{e1}$ (eqn. \ref{e13}). The physical parameters used in the integration are $\mu_1=9.7 \times 10^{-6}$, $\mu_2=4.8\times 10^{-5}$, $m_{star}=0.31 M_{\sun}$, $\tau_{n1}=2 \times 10^5$~years, $\tau_{n2}=4 \times 10^4$~years, $\tau_{n1}/\tau_{e1}=\tau_{n2}/\tau_{e2}=100$.}
\label{figA1}
\end{figure}

\begin{figure} [htp]
\centerline{\epsfig{file=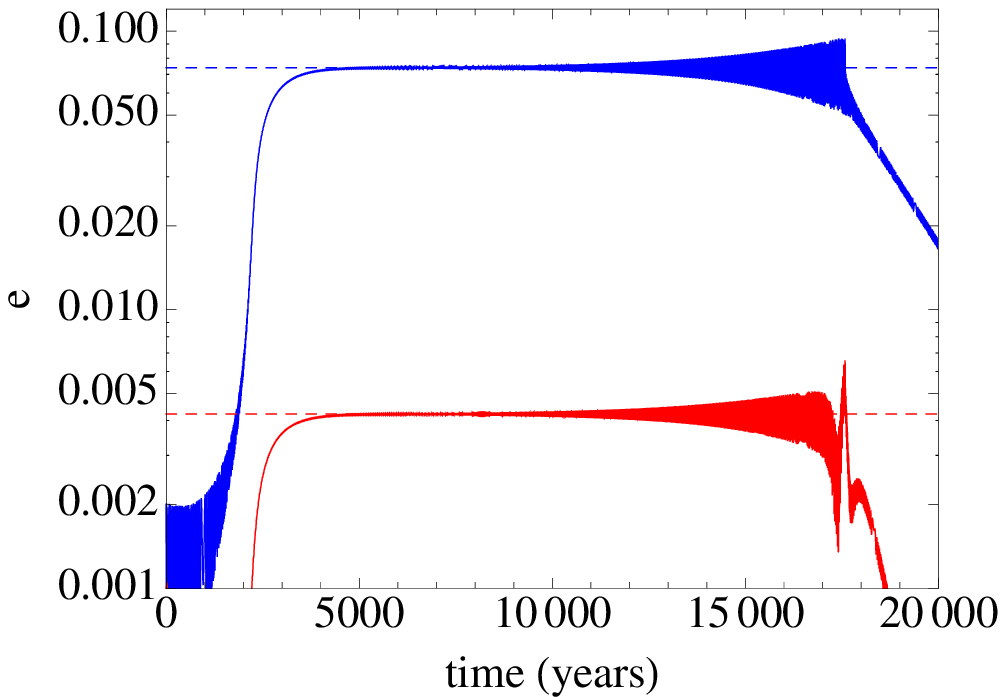, scale=1.4}}
\caption{Eccentricity as a function of time for planet 1 (blue) and planet 2 (red). The solid lines are from integrations of the resonant equations of motion and the dashed lines correspond to the analytic results given in equations (\ref{e59}) and (\ref{e54}) . The physical parameters used in the integration are the same as in Figure \ref{figA1}.}
\label{figA2}
\end{figure}

If the planets are caught in an exact 2:1 mean motion resonance, $2n_2-n_1=0$ and $\dot\varpi_1=\dot\varpi_2$ which implies
\begin{equation}\label{e59}
\frac{e_1}{e_2}= 2\frac{\alpha b}{c}\frac{\mu_2}{\mu_1}\, .
\end{equation} 
Furthermore, in the presence of eccentricity damping, the librations of the resonant argument are no longer centered exactly at $\phi_1=0$ and $\phi_2=\pi$ but are offset from 0 and $\pi$ such that
\begin{equation}
\sin \phi_1 = \frac{e_1}{\alpha b\tau_{e1} \mu_2 n_1},\ \quad \sin \phi_2 = -\frac{e_2}{c\tau_{e2} \mu_1 n_2}\, .
\end{equation}
Assuming that $\tau_{e1}/\tau_{e2} = \tau_{n1}/\tau_{n2}= \mu_2/\mu_1$, as expected for dissipation due to interactions with a protoplanetary disk (see equations (\ref{e17}) and (\ref{e18})), the condition $2n_2-n_1=0$ implies
\begin{equation}\label{e54}
 e_1= \left( \frac{\tau_{e1}}{6\tau_{n2}}\right)^{1/2} \left( \frac{1-\mu_1/\mu_2}{(1+\mu_1/(2\alpha\mu_2))(1+(c/b)^2/(4\alpha))}\right)^{1/2}\, .
\end{equation}
Figures \ref{figA1} and \ref{figA2} display the results of the numerical integration of equations (\ref{e200})-(\ref{e201}).  Figure \ref{figA1} shows the evolution of the semi-major axis of the two planets. Convergent migration leads to temporary capture into 2:1 mean motion resonance, but escape from this resonance occurs after a few eccentricity damping timescales. Figure \ref{figA2} displays the corresponding eccentricity evolution of the two planets (solid lines) and the expected equilibrium eccentricity calculated analytically in equations  (\ref{e59}) and (\ref{e54}) (dashed lines). The numerical and analytical results are in very good agreement.

In the limit that $\mu_1/\mu_2 << 1$, equation (\ref{e54}) simplifies to
 \begin{equation}
 e_1= \left( \frac{\tau_{e1}}{6\tau_{n2}}\right)^{1/2} \left( \frac{1}{(1+(c/b)^2/(4\alpha))}\right)^{1/2} \simeq  0.98 \left( \frac{\tau_{e1}}{6\tau_{n2}}\right)^{1/2}
  \end{equation}
and the two-planet resonance problem reduces to a one-planet resonance problem in which the inner planet migrates toward the outer one at the rate $\tau_{n2}$ such that  
 \begin{equation}
 \dot n_1 -2 \dot n_2 = 3\alpha b \mu_2n_1^2e_1\sin\phi_1-\frac{n_1}{\tau_{n2}}+\frac{3n_1e_1^2}{\tau_{e1}}.
\end{equation}
The right hand side of above equation is identical to equation (\ref{e6}) except the migration rate is that of the outer planet rather than that of the inner one. The equations of motion for $\dot e_1$ and $\dot \varpi_1$ remain unchanged as given by equations (\ref{e46}) and (\ref{e47}), respectively. Taking the ratio of equations (\ref{e17}) and (\ref{e18}) yields
\begin{equation}
\frac{\tau_{e1}}{\tau_{n2}} \sim \left(\frac{h}{a}\right)^2 \left(\frac{\mu_2}{\mu_1} \right),
\end{equation}
which evaluates to about $0.01 (\mu_2/\mu_1)$ for typical disk scale heights $h/a\approx 0.1$. The results of this section demonstrate that a simplified treatment of mean motion resonance based on a single active planet is capable of revealing the essential features of the more complex two planet dynamics provided $\tau_n$ is set equal to $\tau_{n_2}$ and $\tau_e$ is set to $\tau_{e_1}$.  

\subsection{Comparison with Results of other Numerical Integrations}
The claim that the amplitude of librations about $\phi_{eq}\, ,\, e_{eq}$ grows for $e_{eq}> (\beta\mu'/3j)^{1/3}$ is at
the heart of our investigation.  It is based on the analysis of a simple model in which only resonant terms in the disturbing function of lowest order in $e$ are retained.  Naturally we were interested in finding out whether the overstability had shown up in previous work. 

To the best of our knowledge, the overstability was first discovered numerically by \citet{MW08} when studying the tidal evolution of Saturn's satellites. Their numerical results of the parameter range that leads to damping, finite amplitude growth, and escape from resonance are in agreement with our analytic expressions.The fact that \cite{LP02} did not find any overstable librations for the GJ 876 planets is consistent with our analytic findings, since the roughly Jupiter mass planets and the range of values of $\tau_e/\tau_n$ considered in their study should result in damping of librations and permanent resonance capture. The equilibrium eccentricities that they find are somewhat larger than predicted by our analytical results because higher than first order eccentricity terms are required to model the large observed eccentricities (i.e, e ? 0.255) of the GJ 876 system (see also section 2 of \cite{LP02}). \cite{DLCB12} perform numerical simulations of planets caught in 2:1 mean motion resonance but stop planet migration after $10^4$ years to model the disappearance of the disk. For the parameters used in their simulations, our analytic expressions predict overstable librations leading to passage through resonance. Our own integrations predict that  \cite{DLCB12} would have found escape from resonance if they had extended 
the migration time for at least an additional 5000 years.

\section{Comparison with Kepler Multi-Planet Systems}\label{s33}
Our investigation implies that pairs of Kepler planets found close to resonance are comprised of those that were permanently captured and others that were temporarily captured but were caught in resonance when the protoplanetary disk dispersed.  Given plausible ranges of $\mu$ and $\tau_e/\tau_n$, it is not surprising that these total only several percent of the pairs in multi-planet systems.  This argument rests on the assumption that semi-major axes are largely frozen during and subsequent to disk dispersal.

\begin{figure} [htp]
\centerline{\epsfig{file=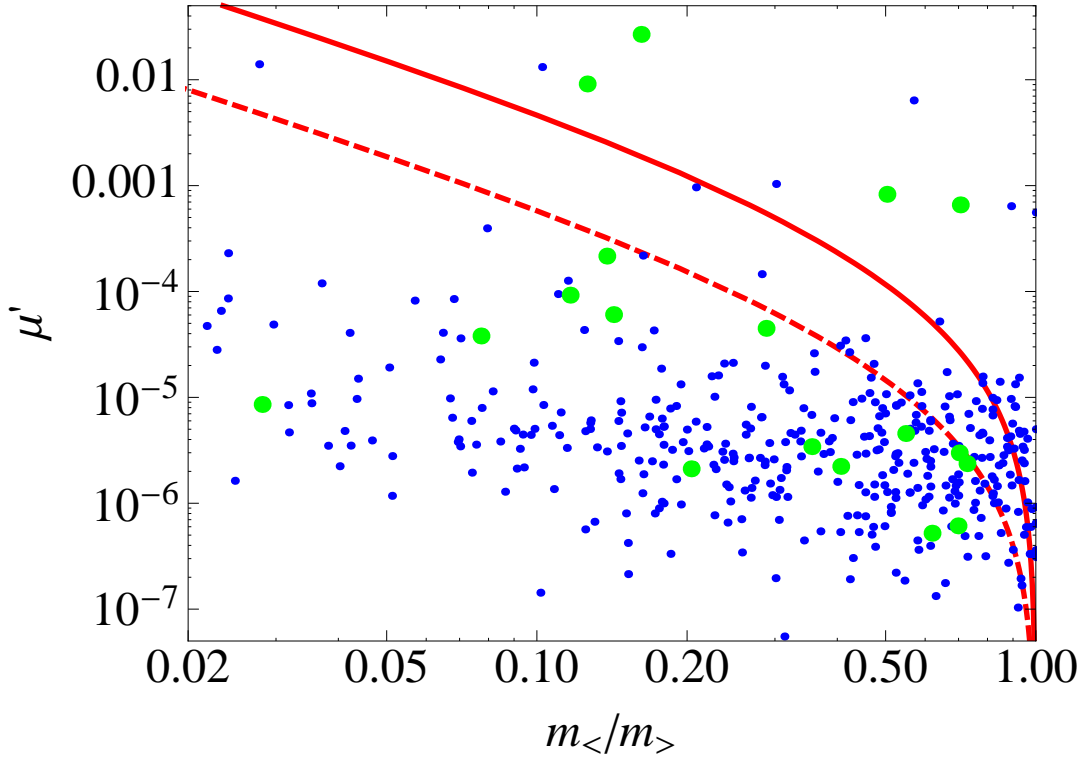, scale=1.4}}
\caption{Planet to Star mass ratio, $\mu'$, as a function of the smaller to larger planet mass ratio, $m_</m_>$, for all Kepler 2-planet systems known as of June 2013 (blue points). 
Large green filled circles correspond to planet pairs with period ratios ranging from 1.99 to 2.075. See Figure 1 for comparison.
Planet pairs above the red dashed line can be permanently captured into resonance.  The solid red line marks the transition from damping to growth for a 2:1 mean motion resonance as given by equation (\ref{e54}) assuming $\tau_{e_<}/\tau_{n_<}=\tau_{e_>}/\tau_{n_>}=0.01$.  Pairs located above the solid red line would have damped resonant arguments (9\%), whereas resonant arguments of those between the dashed and red lines should have undergone finite amplitude librations (20\%). 
The vast majority of pairs might have been temporarily captured in resonances but would have subsequently escaped. Even many pairs located between the dashed and solid lines might have only been captured temporarily as we found for systems with short eccentricity damping timescales, $\tau_{e_<}$, in Figure \ref{fig7}.}\label{fig8}
\end{figure}

\begin{figure} [htp]
\centerline{\epsfig{file=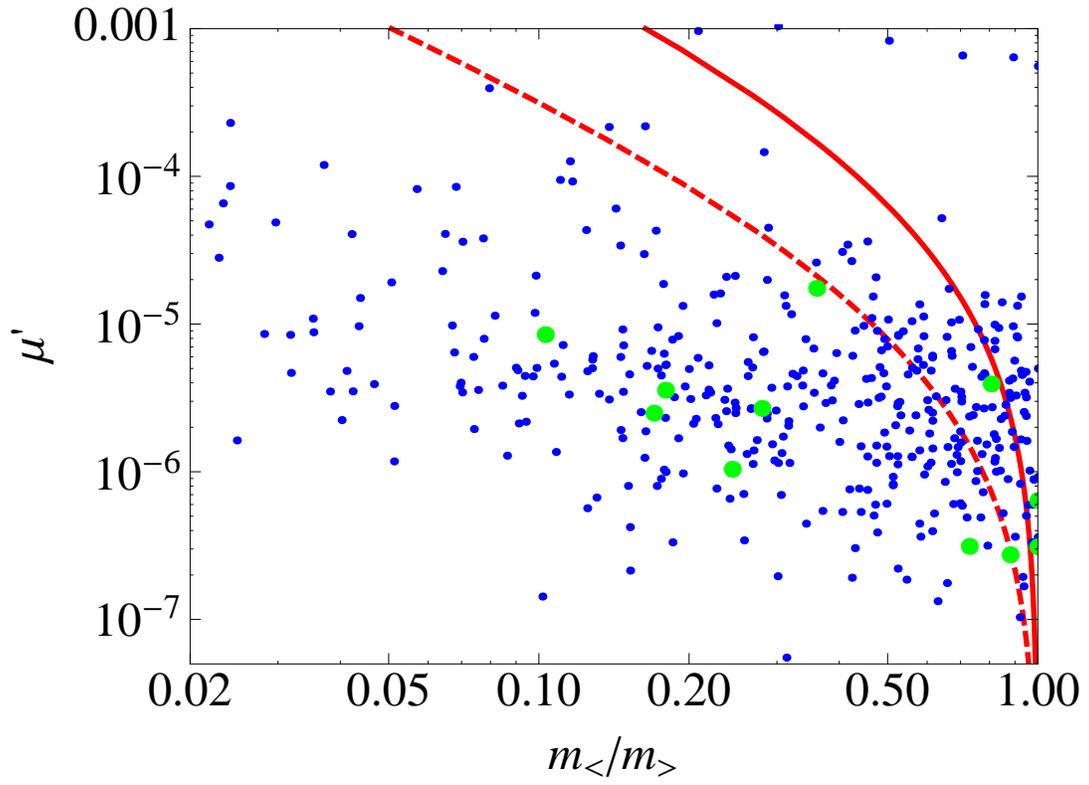, scale=1.4}}
\caption{Same as for Figure \ref{fig8} but for the 3:2 mean motion resonance. Large filled green circles correspond to planet pairs with period ratios ranging from 1.5 to 1.55. See Figure 1 for comparison.}\label{fig82}
\end{figure}

Figures \ref{fig8} and \ref{fig82} plot the more massive planet to star mass ratio, $\mu'$, as a function of the planet pair mass ratio, $m_{<}/m_{>}$, for all Kepler 2-planet systems known as of June 2013. The subscripts $<$ and $>$ refer to the smaller and larger of the two planets, respectively. These ratios were calculated based on the assumption that each planet has a mean density of $2~\rm{g\,cm^{-3}}$.  Planet pairs above the dashed red lines are candidates for  permanent capture in the $2:1$ (fig. \ref{fig8}) and $3:2$ (fig. \ref{fig82}) resonance.  Resonant arguments for those above the solid red lines would be damped whereas those located below it would be overstable.  Most of the planet pairs lie below the dashed red lines.  These might have undergone temporary capture but would have escaped from resonance on timescale $\tau_{e_<}$.  The fate of the pairs between the dashed and solid lines is less certain.  For many, the eccentricity damping timescale, $\tau_{e_<}$, is so short that escape would be probable as found in our numerical integrations summarized in Figure \ref{fig7}. Given that planet pairs which ultimately escape resonance only spend a time $\tau_{e_<}$ in each resonance but about  a time $\tau_{n_>}$ between successive resonances, the fraction temporarily captured planet pairs in or near mean motion resonances is $\tau_{e_<}/\tau_{n_>} \sim 0.01 \mu_>/\mu_<$ which evaluates to 3\% for the Kepler 2-planet sample.  Perhaps a comparable or even slightly greater percent might have been permanently captured.

\section{Departure from Exact Resonance}
\begin{figure} [htp]
\centerline{\epsfig{file=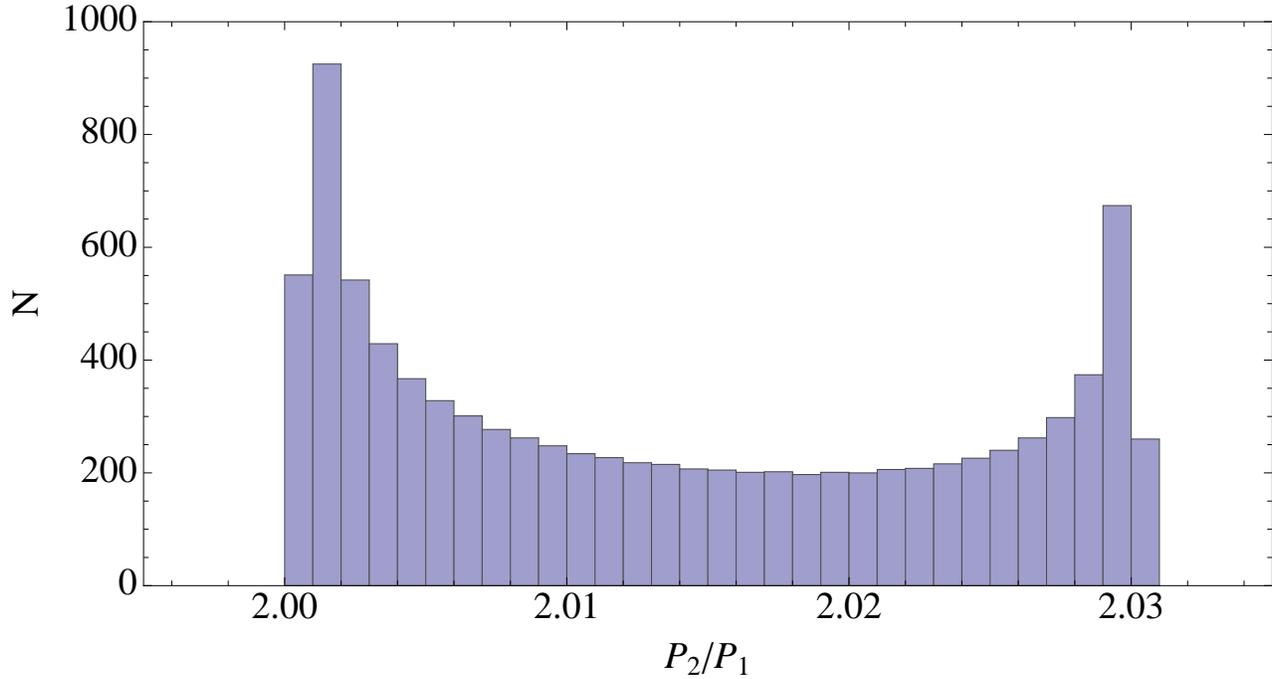, scale=1.1}}
\caption{Histogram of the period distribution for a planet pair permanently trapped in the 2:1 resonance and undergoing a large amplitude libration consistent with parameter values $\mu_1=0$, $\mu_2= 4 \times 10^{-4}$,  and $\tau_{e1}/\tau_{n2} =0.022$.  Period ratios exceed the precise resonance value of $2$ by up to 1.5\%.  The intrinsic asymmetry about exact resonance is responsible for the deficit of period ratios short of 2:1.  However,  as seen from Figure \ref{fig8}, only a few Kepler pairs near the 2:1 resonance have $\mu_2$ as large as $4\times 10^{-4}$. Thus it appears that additional sources of eccentricity damping must have occurred during or after the disk disappeared. }
\label{fig6}
\end{figure}

Figure \ref{fig1} shows that peaks associated with period ratios close to $2:1$ and $3:2$ are systematically displaced to larger values by of order $1-2\%$.  The direction of displacement is almost certainly due to the asymmetry that requires convergent migration for capture in resonance.  However, its magnitude is less easily accounted for.  This is
illustrated in Figure \ref{fig6} which shows that to match a $1\%$ average offset from exact $2:1$ resonance requires a permanently captured test particle to be paired with a
planet for which $\mu_2\sim 4\times 10^{-4}$.  More generally, test particles paired in mean motion resonances with massive planets would have period ratios offset from exact
resonance by $\sim j^{1/3}\mu_2^{2/3}$ (cf. eqn. \ref{e102}).  \citet{PMT13} arrive at similar mass requirements and scalings from a model in which planets grow in mass at a prescribed rate without orbital migration or eccentricity damping.
Period offsets for planets temporarily caught in resonance that might remain close to resonance after the disk
dissipates would be smaller still.  These considerations strongly suggest that the departure from exact resonance observed in the Kepler data occurred either during or after the
protoplanetary disk disappeared.  

Retreat from resonance could occur as a result of eccentricity damping provided it were slow on the timescale of small amplitude librations about the fixed point at $\phi=0, \, e=e_0$.  Several mechanisms come to mind.  Prominent among them are tides raised by the star in the planets as suggested by \cite{LW12} and \cite{BM13}, dynamical friction due to a residual particle disk, and the excitation of apsidal waves.  Proposals involving tidal damping of eccentricity \citep{LW12,BM13} have been critically examined by \cite{LFL13} who conclude that with reasonable estimates of the tidal Love number, $k$, and dissipation factor, $Q$, tidal damping cannot account for the majority of offsets from resonance.  Recent work by \citet{BP13} finds that interactions between a planet and the wake of a companion can reverse convergent migration and significantly increase the period ratio away from the resonant value. They suggest that this may help to account for the diversity of period ratios in KeplerÕs multiple planet systems.

If a planetesimal disk contains a significant mass, interactions between the planets and the disk may lead to significant damping of orbital eccentricities. The resonant structure of the Kuiper Belt and the 'late veneer' found on Earth, Moon and Mars provide evidence for a leftover planetesimal disk in our solar system that was still present at the end of planet formation \citep{SWY12}. 

In some instances,  interactions with a particle disk can be considered in analogy to those with a gaseous disk.  If the disk were composed of small bodies, collisions might maintain the ratio of its thickness to radius at a much lower value than typical of a gaseous disk.  This could lower $\tau_e/\tau_n$ and perhaps provide sufficient eccentricity damping to match the observed offsets from exact resonance.  However, this proposal is plagued by uncertainty.  Gaps are easily opened in disks with small velocity dispersion and the absence of material corotating with the planet would eliminate eccentricity damping associated with coorbital Lindblad resonances. Coorbital Lindblad resonances dominate eccentricity damping during type I migration. 

By virtue of their slow detuning away from exact resonance, apsidal waves launched at a secular resonance can propagate in low optical depth disks where waves associated with other Lindblad resonances cannot \citep{GT78, WH98}.  That an apsidal wave transports negative angular momentum follows from the fact that it excites the eccentricities of disk particles while hardly affecting their energies. As a consequence, the excitation of the apsidal wave damps the
planet's orbital eccentricity.  There is no difficulty conjuring up reasonable scenarios in which the apsidal wave is responsible for damping a planet's orbital eccentricity by a few orders of magnitude within the timescale that the 
particle disk might survive destruction or accretion onto planets and the central star.  However, just as for damping by dynamical friction, uncertainties abound.  Foremost among them are the mass and lifetime of the particle disk, the particle size distribution, and the radial dependence of the apsidal precession rate.  We leave these issues for a future investigation.

\section{DISCUSSION AND CONCLUSIONS}

Our investigation is predicated on the assumption that the orbits of the Kepler planets were largely determined by processes that operated within their protoplanetary disks and that they have undergone little modification since it disappeared.  That $\tau_e/\tau_n$, which depends on temperature but is independent of disk surface density,  plays a central role in the outcome of evolution in resonance is consistent with, but far from a proof that this assumption is valid. Given this starting point, we argue that the small fraction of Kepler pairs found close to mean motion resonance is compatible with standard estimates for the rates of orbital migration and eccentricity damping due to planet-disk interactions.  Figure \ref{fig3} shows that capture into first-order mean motion resonance during convergent migration must have been a common occurrence given the observed orbital parameters and estimated planet masses.  However, as the result of  eccentricity damping, permanent capture in resonance was rare.  Most captures were only temporary with durations of order $\tau_e$ which is short in comparison to the timescale $\tau_n$ for migration between neighboring resonances.  The temporary nature of resonance capture is due to the overstability of librations of the resonant argument about the tidal equilibrium given by
equation (\ref{e106}).  The paucity of resonances among Kepler pairs should therefore not be taken as evidence for in situ planet formation or the disruptive effects of disk turbulence \citep{R12}.  

The resonance model and our analytic solutions presented here only include the lowest order term in eccentricity. The assumption of small eccentricities seems to be justified for the majority of Kepler systems given that estimates for typical disk parameters yield equilibrium eccentricities smaller than 0.1.

The overstability criterion described in \S \ref{ssoverstab} is the major technical accomplishment of our paper.  Permanent resonance capture is possible if tidal equilibrium occurs at large enough separation from resonance so that a separatrix is not present.  As indicated by Figures \ref{fig8} and \ref{fig82}, only the most massive Kepler planets are likely to be permanently captured in resonance.  For the others,  overstable librations lead to passage through resonance and to the resumption of unimpeded convergent migration.  In this context, the drop off in the number of pairs with period ratios below $1.3$ seen in Figure \ref{fig1} presents a puzzle, but it may, at least partly, be due to chaos caused by resonance overlap \citep{W80,DPH13}.

\acknowledgements{HS gratefully acknowledges support from the Wade Fund. We thank Scott Tremaine, Jing Luan and Glen Stewart for useful comments that helped to improve this manuscript.}

\appendix
\section{Appendix}
Equilibrium eccentricities and conditions for overstable librations were derived in the body of our paper under the assumption that eccentricity is damped at constant angular momentum such that p=3 in equation (\ref{e6}). Here we relax this assumption and provide expressions for the equilibrium eccentricity and conditions for overstable librations appropriate for a general value of $p>0$. Analogous expressions to the ones derived in section \ref{ssedamp} are as follows:  Equation ({\ref{e106}) for the equilibrium eccentricity generalizes to 
\begin{equation}\label{e706}
e_{eq}= \left( \frac{\tau_{e}}{(3j+p)\tau_{n}}\right)^{1/2}.
\end{equation} 
Equation (\ref{e13}) for the growth rate of overstability is replaced by
\begin{equation}\label{e713}
s=\frac{1}{\tau_{e}}\left(\frac{n}{\omega_{eq}}\right)^2\left(jp\beta \mu'e_{eq}-\left(\frac{\beta\mu'}{e_{eq}}\right)^2\right)\, .
\end{equation}
Thus overstable librations require $p > 0$. Equations (\ref{e706}) and (\ref{e713}) imply that a planet is permanently trapped in resonance and its librations damped provided
\begin{equation}\label{e708}
\mu' > \frac{jp}{(3j+p)^{3/2}\beta}\left(\tau_{e}\over \tau_{n}\right)^{3/2}\,.
\end{equation}
For 
\begin{equation}\label{e709}
\frac{j^2p}{8(3j+p)^{3/2}\beta}\left(\tau_{e}\over \tau_{n}\right)^{3/2} < \mu' < \frac{jp}{(3j+p)^{3/2}\beta}\left(\tau_{e}\over \tau_{n}\right)^{3/2}\, ,
\end{equation}
the planet is permanently caught in resonance and its libration amplitude saturates at a finite value. Lastly, for 
\begin{equation}\label{e710}
\mu' < \frac{j^2p}{8(3j+p)^{3/2}\beta}\left(\tau_{e}\over \tau_{n}\right)^{3/2}\, ,
\end{equation}
the planet is caught in resonance but then escapes on timescale $\tau_e$.


\begin{thebibliography}{30}
\expandafter\ifx\csname natexlab\endcsname\relax\def\natexlab#1{#1}\fi
\expandafter\ifx\csname url\endcsname\relax
  \def\url#1{\texttt{#1}}\fi
\expandafter\ifx\csname urlprefix\endcsname\relax\def\urlprefix{URL }\fi
\providecommand{\eprint}[2][]{\url{#2}}

\bibitem[{{Artymowicz}(1993)}]{Artymowicz93}
{Artymowicz}, P. 1993, \apj, 419, 166

\bibitem[{{Baruteau} \& {Papaloizou}(2013)}]{BP13}
{Baruteau}, C., \& {Papaloizou}, J.~C.~B. 2013, ArXiv e-prints.
  \eprint{1301.0779}

\bibitem[{{Batalha} et~al.(2013){Batalha}, {Rowe}, {Bryson}, {Barclay},
  {Burke}, {Caldwell}, {Christiansen}, {Mullally}, {Thompson}, {Brown},
  {Dupree}, {Fabrycky}, {Ford}, {Fortney}, {Gilliland}, {Isaacson}, {Latham},
  {Marcy}, {Quinn}, {Ragozzine}, {Shporer}, {Borucki}, {Ciardi}, {Gautier},
  {Haas}, {Jenkins}, {Koch}, {Lissauer}, {Rapin}, {Basri}, {Boss}, {Buchhave},
  {Carter}, {Charbonneau}, {Christensen-Dalsgaard}, {Clarke}, {Cochran},
  {Demory}, {Desert}, {Devore}, {Doyle}, {Esquerdo}, {Everett}, {Fressin},
  {Geary}, {Girouard}, {Gould}, {Hall}, {Holman}, {Howard}, {Howell},
  {Ibrahim}, {Kinemuchi}, {Kjeldsen}, {Klaus}, {Li}, {Lucas}, {Meibom},
  {Morris}, {Pr{\v s}a}, {Quintana}, {Sanderfer}, {Sasselov}, {Seader},
  {Smith}, {Steffen}, {Still}, {Stumpe}, {Tarter}, {Tenenbaum}, {Torres},
  {Twicken}, {Uddin}, {Van Cleve}, {Walkowicz}, \& {Welsh}}]{B13}
{Batalha}, N.~M., {Rowe}, J.~F., {Bryson}, S.~T., {Barclay}, T., {Burke},
  C.~J., {Caldwell}, D.~A., {Christiansen}, J.~L., {Mullally}, F., {Thompson},
  S.~E., {Brown}, T.~M., {Dupree}, A.~K., {Fabrycky}, D.~C., {Ford}, E.~B.,
  {Fortney}, J.~J., {Gilliland}, R.~L., {Isaacson}, H., {Latham}, D.~W.,
  {Marcy}, G.~W., {Quinn}, S.~N., {Ragozzine}, D., {Shporer}, A., {Borucki},
  W.~J., {Ciardi}, D.~R., {Gautier}, T.~N., III, {Haas}, M.~R., {Jenkins},
  J.~M., {Koch}, D.~G., {Lissauer}, J.~J., {Rapin}, W., {Basri}, G.~S., {Boss},
  A.~P., {Buchhave}, L.~A., {Carter}, J.~A., {Charbonneau}, D.,
  {Christensen-Dalsgaard}, J., {Clarke}, B.~D., {Cochran}, W.~D., {Demory},
  B.-O., {Desert}, J.-M., {Devore}, E., {Doyle}, L.~R., {Esquerdo}, G.~A.,
  {Everett}, M., {Fressin}, F., {Geary}, J.~C., {Girouard}, F.~R., {Gould}, A.,
  {Hall}, J.~R., {Holman}, M.~J., {Howard}, A.~W., {Howell}, S.~B., {Ibrahim},
  K.~A., {Kinemuchi}, K., {Kjeldsen}, H., {Klaus}, T.~C., {Li}, J., {Lucas},
  P.~W., {Meibom}, S., {Morris}, R.~L., {Pr{\v s}a}, A., {Quintana}, E.,
  {Sanderfer}, D.~T., {Sasselov}, D., {Seader}, S.~E., {Smith}, J.~C.,
  {Steffen}, J.~H., {Still}, M., {Stumpe}, M.~C., {Tarter}, J.~C., {Tenenbaum},
  P., {Torres}, G., {Twicken}, J.~D., {Uddin}, K., {Van Cleve}, J.,
  {Walkowicz}, L., \& {Welsh}, W.~F. 2013, \apjs, 204, 24. \eprint{1202.5852}

\bibitem[{{Batygin} \& {Morbidelli}(2013)}]{BM13}
{Batygin}, K., \& {Morbidelli}, A. 2013, \aj, 145, 1. \eprint{1204.2791}

\bibitem[{{Borderies} \& {Goldreich}(1984)}]{BG84}
{Borderies}, N., \& {Goldreich}, P. 1984, Celestial Mechanics, 32, 127

\bibitem[{{Deck} et~al.(2013){Deck}, {Payne}, \& {Holman}}]{DPH13}
{Deck}, K.~M., {Payne}, M., \& {Holman}, M.~J. 2013, \apj, 774, 129.
  \eprint{1307.8119}

\bibitem[{{Delisle} et~al.(2012){Delisle}, {Laskar}, {Correia}, \&
  {Bou{\'e}}}]{DLCB12}
{Delisle}, J.-B., {Laskar}, J., {Correia}, A.~C.~M., \& {Bou{\'e}}, G. 2012,
  \aap, 546, A71. \eprint{1207.3171}

\bibitem[{{Fabrycky} et~al.(2012){Fabrycky}, {Lissauer}, {Ragozzine}, {Rowe},
  {Agol}, {Barclay}, {Batalha}, {Borucki}, {Ciardi}, {Ford}, {Geary}, {Holman},
  {Jenkins}, {Li}, {Morehead}, {Shporer}, {Smith}, {Steffen}, \&
  {Still}}]{FB12}
{Fabrycky}, D.~C., {Lissauer}, J.~J., {Ragozzine}, D., {Rowe}, J.~F., {Agol},
  E., {Barclay}, T., {Batalha}, N., {Borucki}, W., {Ciardi}, D.~R., {Ford},
  E.~B., {Geary}, J.~C., {Holman}, M.~J., {Jenkins}, J.~M., {Li}, J.,
  {Morehead}, R.~C., {Shporer}, A., {Smith}, J.~C., {Steffen}, J.~H., \&
  {Still}, M. 2012, ArXiv e-prints. \eprint{1202.6328}

\bibitem[{{Friedland}(2001)}]{Friedland01}
{Friedland}, L. 2001, \apjl, 547, L75

\bibitem[{{Goldreich}(1965)}]{Goldreich65}
{Goldreich}, P. 1965, \mnras, 130, 159

\bibitem[{{Goldreich} \& {Tremaine}(1980)}]{GT80}
{Goldreich}, P., \& {Tremaine}, S. 1980, \apj, 241, 425

\bibitem[{{Goldreich} \& {Tremaine}(1978)}]{GT78}
{Goldreich}, P., \& {Tremaine}, S.~D. 1978, \icarus, 34, 240

\bibitem[{{Henrard}(1982)}]{Henrard82}
{Henrard}, J. 1982, Celestial Mechanics, 27, 3

\bibitem[{{Lee} et~al.(2013){Lee}, {Fabrycky}, \& {Lin}}]{LFL13}
{Lee}, M.~H., {Fabrycky}, D., \& {Lin}, D.~N.~C. 2013, ArXiv e-prints.
  \eprint{1307.4874}

\bibitem[{{Lee} \& {Peale}(2002)}]{LP02}
{Lee}, M.~H., \& {Peale}, S.~J. 2002, \apj, 567, 596

\bibitem[{{Lithwick} \& {Wu}(2012)}]{LW12}
{Lithwick}, Y., \& {Wu}, Y. 2012, \apjl, 756, L11. \eprint{1204.2555}

\bibitem[{{Malhotra}(1993)}]{M93}
{Malhotra}, R. 1993, \nat, 365, 819

\bibitem[{{Meyer} \& {Wisdom}(2008)}]{MW08}
{Meyer}, J., \& {Wisdom}, J. 2008, \icarus, 193, 213

\bibitem[{{Murray} \& {Dermott}(1999)}]{MD99}
{Murray}, C.~D., \& {Dermott}, S.~F. 1999, {Solar system dynamics}

\bibitem[{{Ogihara} \& {Kobayashi}(2013)}]{OK13}
{Ogihara}, M., \& {Kobayashi}, H. 2013, ArXiv e-prints. \eprint{1307.7776}

\bibitem[{{Petrovich} et~al.(2013){Petrovich}, {Malhotra}, \&
  {Tremaine}}]{PMT13}
{Petrovich}, C., {Malhotra}, R., \& {Tremaine}, S. 2013, \apj, 770, 24.
  \eprint{1211.5603}

\bibitem[{{Quillen}(2006)}]{Quillen06}
{Quillen}, A.~C. 2006, \mnras, 365, 1367. \eprint{arXiv:astro-ph/0507477}

\bibitem[{{Rein}(2012)}]{R12}
{Rein}, H. 2012, \mnras, 427, L21. \eprint{1208.3583}

\bibitem[{{Roy} \& {Ovenden}(1954)}]{RO54}
{Roy}, A.~E., \& {Ovenden}, M.~W. 1954, \mnras, 114, 232

\bibitem[{{Schlichting} et~al.(2012){Schlichting}, {Warren}, \& {Yin}}]{SWY12}
{Schlichting}, H.~E., {Warren}, P.~H., \& {Yin}, Q.-Z. 2012, \apj, 752, 8.
  \eprint{1202.6372}
  
\bibitem[{{Tanaka} \& {Ward}(2004)}]{TW04}
{Tanaka}, H., \& {Ward}, W.~R. 2004, \apj, 602, 388

\bibitem[{{Ward}(1986)}]{Ward86}
{Ward}, W.~R. 1986, \icarus, 67, 164

\bibitem[{{Ward}(1988)}]{Ward88}
--- 1988, \icarus, 73, 330

\bibitem[{{Ward} \& {Hahn}(1998)}]{WH98}
{Ward}, W.~R., \& {Hahn}, J.~M. 1998, \aj, 116, 489

\bibitem[{{Wisdom}(1980)}]{W80}
{Wisdom}, J. 1980, \aj, 85, 1122

\bibitem[{{Yoder}(1979)}]{Yoder79}
{Yoder}, C.~F. 1979, Celestial Mechanics, 19, 3

\end{thebibliography}

\end{document}